\documentclass[twocolumn,english,showpacs,preprintnumbers,jap]{revtex4-1}
\usepackage[latin9]{inputenc}
\usepackage{amsmath}
\usepackage{amssymb}
\usepackage{graphicx}
\usepackage{esint}
\usepackage{color}
\usepackage{ulem}
\makeatletter
\@ifundefined{textcolor}{}
{%
 \definecolor{BLACK}{gray}{0}
 \definecolor{WHITE}{gray}{1}
 \definecolor{RED}{rgb}{1,0,0}
 \definecolor{GREEN}{rgb}{0,1,0}
 \definecolor{BLUE}{rgb}{0,0,1}
 \definecolor{CYAN}{cmyk}{1,0,0,0}
 \definecolor{MAGENTA}{cmyk}{0,1,0,0}
 \definecolor{YELLOW}{cmyk}{0,0,1,0}
 \definecolor{red}{rgb}{1,0,0}

\definecolor{blue}{rgb}{0,0,1}
}

\usepackage{babel}

\makeatother

\usepackage{babel}
\begin{document}

\title{Probing the Antisymmetric Fano Interference Assisted by a Majorana
Fermion}

\author{F. A. Dessotti$^{1}$, L. S. Ricco$^{1}$, M. de Souza$^{2},$ F.
M. Souza$^{3}$, and A. C. Seridonio$^{1,2}$ }

\affiliation{$^{1}$Departamento de F\'{i}sica e Qu\'{i}mica, Unesp - Univ Estadual
Paulista, 15385-000, Ilha Solteira, SP, Brazil\\
$^{2}$IGCE, Unesp - Univ Estadual Paulista, Departamento de F\'{i}sica, 13506-900, Rio Claro, SP, Brazil\\
$^{3}$Instituto de F\'{i}sica, Universidade Federal de Uberlândia,
38400-902, Uberlândia, MG, Brazil}
\begin{abstract}
As the Fano effect is an interference phenomenon where tunneling paths compete for the electronic transport, it becomes a probe to catch fingerprints of  Majorana fermions lying on condensed matter systems. In this work we benefit of this mechanism by proposing as a route for that an Aharonov-Bohm-like interferometer composed by
two quantum dots, being one of them coupled to a Majorana bound state, which is attached
to one of the edges of a semi-infinite Kitaev wire within the topological phase. By changing the Fermi energy of the leads and the symmetric
detuning of the levels for the dots, we show that opposing Fano regimes result in a transmittance
characterized by distinct conducting and insulating regions, which
are fingerprints of an isolated Majorana quasiparticle. Furthermore,
we show that the maximum fluctuation of the transmittance as a function of the detuning
is half for a semi-infinite wire, while it corresponds to the unity for a finite system. The setup proposed here constitutes an alternative experimental tool to detect Majorana excitations.
\end{abstract}


\maketitle

\section{Introduction}
\label{sec1}

In the field of high-energy Physics, Ettore Majorana proposed, almost a century ago,
the existence of fermions that form their own antiparticles. In
condensed matter Physics, such fermions emerge as quasiparticle excitations \cite{key-24}. Remarkably, two distant Majorana quasiparticles can
define a single nonlocal regular fermion, which provides a protected
qubit, free of the environment and immune to the decoherence phenomenon.
Such a bit thus can be considered as the fundamental unity for the achievement
of a quantum computer. For this reason, in the last decade, the run for
devices based on Majoranas started in the field of quantum information
\cite{key-36,key-37}. In this scenario, the superconducting state
is the most promising environment for the feasibility of Majorana fermions, in
particular, the $p$-wave and spinless type.

The Kitaev wire within
the topological phase \cite{key-333} is an example, since Majorana fermions
emerge as zero-energy modes bounded to the edges of such a system.
From the experimental point of view, the realization of $p$-wave
superconductivity can be performed by putting an $s$-wave superconductor
close to a semiconducting nanowire with strong spin-orbit interaction
placed perpendicular to a huge magnetic field. In such conditions,
$p$-wave superconductivity is induced on the nanowire due to
the so-called proximity effect \cite{key-DLoss1,key-DLoss2,key-PSodano1}.
\begin{figure}
\includegraphics[width=0.46\textwidth,height=0.2\textheight]{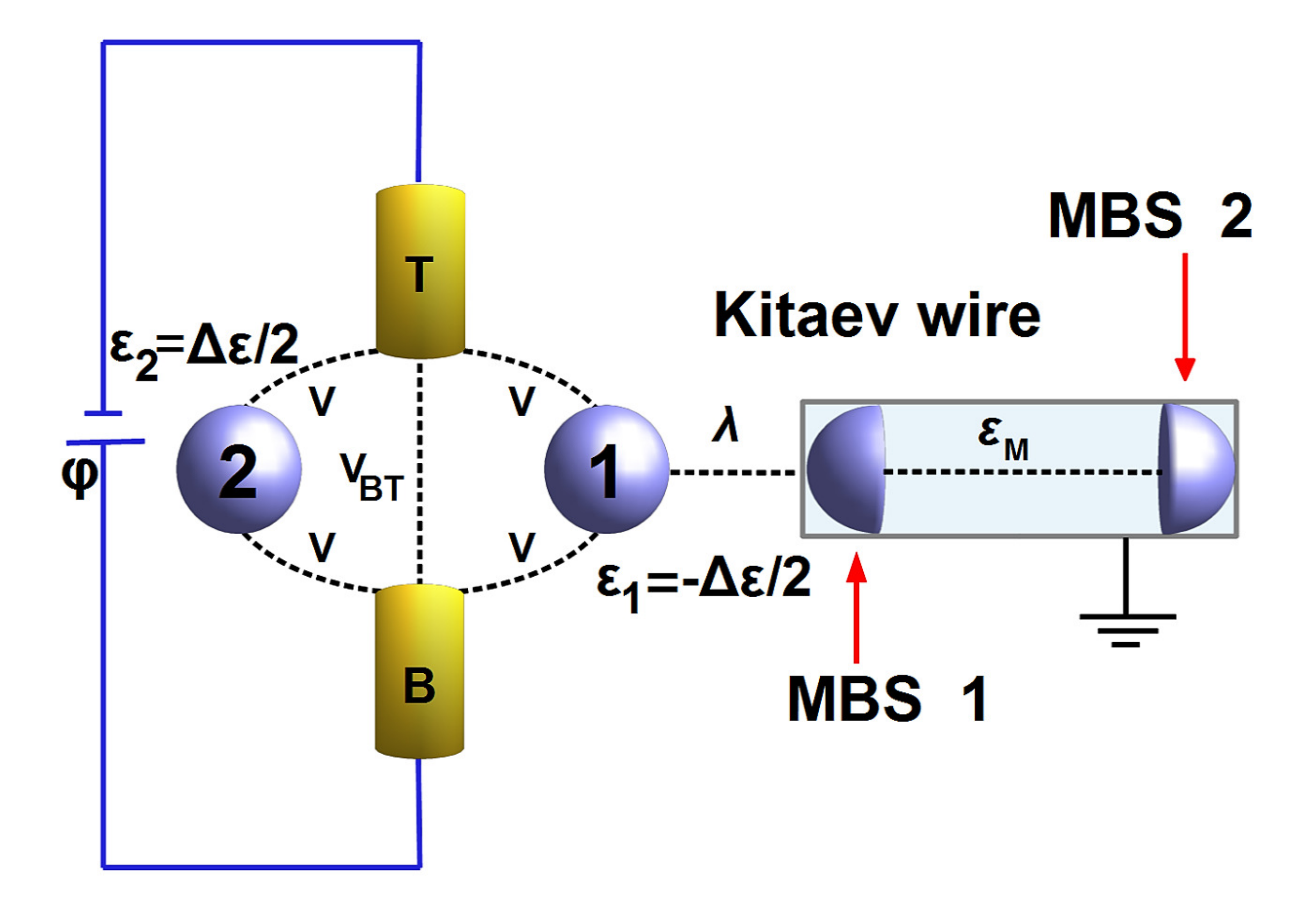}
\caption{\label{Fig1}(Color online) Aharonov-Bohm-like interferometer with
MBSs hosted by a Kitaev wire within the topological phase: two leads
are coupled to two QDs hybridized with a MBS 1; the half-electron
state is represented by the semi-sphere in the left side of the wire.
The tunneling parameters between the QDs and the leads are $V$ and the lead-lead
coupling is $V_{BT}$. The symmetric detuning in the QDs is $\Delta\varepsilon=\varepsilon_{2}-\varepsilon_{1},$
where $\varepsilon_{1}=-\frac{\Delta\varepsilon}{2}$ and $\varepsilon_{2}=\frac{\Delta\varepsilon}{2}$
represent the energy levels of the QDs. The coupling strength of the QD
1 with the MBS 1 is $\lambda$. $\varepsilon_{M}$ couples the MBS
1 to the MBS 2 (the semi-sphere in the right side of the wire). The
bias $\varphi$ of the setup approaches zero.}
\end{figure}

In the context of quantum transport \cite{key-371,JAP1,JAP2,key-25,key-25b,key-25c,Ueda,key-374,key-301,key-302,key-RA1,key-RA2},
in particular for a single quantum dot (QD) hybridized with a Majorana
bound state (MBS), a zero-bias anomaly (ZBA)
\cite{key-25,key-25b} in the conductance is predicted
to appear, given by $G=0.5G_{0}$, where $G_{0}=e^{2}/h$
is the quantum of conductance. The ZBA has been observed experimentally in conductance measurements
through a nanowire of indium antimonide merged
to gold and niobium titanium nitride \cite{key-301}. In this system, Majoranas are supposed to exist as a result of the ZBA that persists
to large magnetic fields and gate voltages. Such a robustness
of the ZBA has also been observed in the setup of a superconductor
of aluminium close to a nanowire of indium arsenide \cite{key-302}.

In this work we follow the strategy of electronic interferometry to probe signatures arising from Majorana fermions. To that end, the Fano interference \cite{Fano1} is the proper way to accomplish the aforementioned goal. As such a phenomenon is due to the competition between  paths of itinerant electrons that travel directly through an energy continuum (an electronic reservoir of a metallic lead for instance) and those that are discrete as found within QDs, patterns of interference then record imprints of Majorana excitations on transmittance profiles. Thereby we explore the manifestation of the Fano effect in the quantum transport through an Aharonov-Bohm-like interferometer formed by two QDs \cite{key-E1,key-E2}, in which one
of them is coupled to a MBS hosted by a semi-infinite Kitaev wire
within the topological phase.

By calculating the transmittance of
this device, we have found that the Fano interference exhibits an
antisymmetric feature: the Green's functions $\tilde{\mathcal{G}}_{d_{j}d_{j}}$
and $\tilde{\mathcal{G}}_{d_{j}d_{l}}$ for the QDs cannot be determined
by the exchange of the indexes $j\leftrightarrow l$ in $\tilde{\mathcal{G}}_{d_{l}d_{l}}$
and $\tilde{\mathcal{G}}_{d_{l}d_{j}}$, respectively, with $l,j=1...2.$
We propose that such a feature can be captured experimentally by performing
measurements of the zero-bias conductance as a function of the Fermi
energy of the leads and the symmetric detuning $\Delta\varepsilon=\varepsilon_{2}-\varepsilon_{1}$
in the QDs, where $\varepsilon_{1}=-\frac{\Delta\varepsilon}{2}$
and $\varepsilon_{2}=\frac{\Delta\varepsilon}{2}$ represent the energy levels
of the QDs \cite{OBS}. As a result, contrasting Fano limits reveal distinct
conducting and insulating regions, which are signatures of an isolated
Majorana quasiparticle.

Moreover, we have determined that the symmetric
Fano effect is recovered when electrons travel only through the
QDs and the Fermi energy of the leads is in resonance with the Majorana
zero mode. As a consequence, the transmittance as a function of the symmetric
detuning exhibits a maximum fluctuation of half for the semi-infinite Kitaev
wire, contrasting with the unity variation observed
in the case of a finite system. In the former situation, the maximum fluctuation of the transmittance is halved due to the half fermion nature of the isolated MBS, while the latter
represents the situation where two MBSs displaced far apart are coupled and
thus forming a nonlocal Dirac fermion delocalized over the wire edges.

This paper is organized as follows: in Sec. \ref{sec2} we develop the theoretical model for the system sketched
in Fig. \ref{Fig1} by deriving the expression for the transmittance through such a device and the Green's functions of the QDs. The results are present in Sec. \ref{sec3} and in Sec.
\ref{sec4}, we summarize our concluding remarks.

\section{The model}
\label{sec2}

In order to mimic the system outlined in Fig.\,\ref{Fig1}, we employ
the Hamiltonian proposed by Liu \textit{et al.} \cite{key-25}, taking
two QDs into account,
\begin{align}
\mathcal{H} & =\sum_{\alpha k}\tilde{\varepsilon}_{\alpha k}c_{\alpha k}^{\dagger}c_{\alpha k}+\sum_{j}\varepsilon_{j}d_{j}^{\dagger}d_{j}+V\sum_{\alpha kj}(c_{\alpha k}^{\dagger}d_{j}+\text{{H.c.}})\nonumber \\
 & +V_{BT}\sum_{kp}(c_{Bk}^{\dagger}c_{Tp}+\text{{H.c.}})+\mathcal{H}_{\text{{MBSs}}},\label{eq:TIAM}
\end{align}
where the electrons in the lead $\alpha=B,T$ (bottom/top) are described
by the operator $c_{\alpha k}^{\dagger}$ ($c_{\alpha k}$) for the
creation (annihilation) of an electron in a quantum state labeled
by the wave number $k$ and energy $\tilde{\varepsilon}_{\alpha k}=\varepsilon_{k}-\mu_{\alpha}$,
with $\mu_{\alpha}$ as the chemical potential. Here we adopt the
gauge $\mu_{B}=\Delta\mu$ and $\mu_{T}=-\Delta\mu$, with $\mu_{B}-\mu_{T}=2\Delta\mu=e\varphi$
as the bias between the leads, being $e>0$ the electron charge and
$\varphi$ the bias-voltage. For the QDs, $d_{j}^{\dagger}$ ($d_{j}$)
creates (annihilates) an electron in the state $\varepsilon_{j}$,
with $j=1,2$. $V$ is the hybridization of the QDs with the leads,
$V_{BT}$ is the lead-lead coupling and
\begin{equation}
\mathcal{H}_{\text{{MBSs}}}=i\varepsilon_{M}\Psi_{1}\Psi_{2}+\lambda(d_{1}-d_{1}^{\dagger})\Psi_{1}\label{eq:MBSs}
\end{equation}
for the Kitaev wire within the topological phase. In particular for
$j=1$, the QD 1 is coupled to the MBS 1 described by the operator
$\Psi_{1}^{\dagger}=\Psi_{1}$. The strength of this coupling is $\lambda$.
The MBS 2 given by $\Psi_{2}^{\dagger}=\Psi_{2}$ is connected to
the MBS 1 via the coefficient $\varepsilon_{M}\sim e^{-L/\xi}$, with
$L$ as the distance between the MBSs and $\xi$ as a coherence length.
In what follows we derive the Landauer-Büttiker formula for the zero-bias conductance
$G$ \cite{book}. Such a quantity is a function of the transmittance $\mathcal{T}\left(\varepsilon\right)$
as follows:
\begin{equation}
G=\frac{e^{2}}{h}\int d\varepsilon\left(-\frac{\partial f_{F}}{\partial\varepsilon}\right)\mathcal{T}(\varepsilon).\label{eq:_10b}
\end{equation}

We begin with the transformations $c_{Bk}=\frac{1}{\sqrt{2}}(c_{ek}+c_{ok})$
and $c_{Tk}=\frac{1}{\sqrt{2}}(c_{ek}-c_{ok})$ on the Hamiltonian
of Eq.\,(\ref{eq:TIAM}), which depends on the \textit{even} and \textit{odd}
conduction operators $c_{ek}$ and $c_{ok}$, respectively. These
definitions allow us to express Eq.\,(\ref{eq:TIAM}) as $\mathcal{H}=\mathcal{H}_{e}+\mathcal{H}_{o}+\mathcal{\tilde{H}}_{\text{{tun}}}=\mathcal{H}_{\varphi=0}+\mathcal{\tilde{H}}_{\text{{tun}}},$
where
\begin{eqnarray}
\mathcal{H}_{e} & = & \sum_{k}\varepsilon_{k}c_{ek}^{\dagger}c_{ek}+\sum_{j}\varepsilon_{j}d_{j}^{\dagger}d_{j}+\sqrt{2}V\sum_{jk}(c_{ek}^{\dagger}d_{j}+\text{{H.c.}})\nonumber \\
 & + & V_{BT}\sum_{kp}c_{ek}^{\dagger}c_{ep}+\mathcal{H}_{\text{{MBSs}}}\label{eq:even}
\end{eqnarray}
represents the Hamiltonian part of the system coupled to the QDs via
an effective hybridization $\sqrt{2}V$, while
\begin{equation}
\mathcal{H}_{o}=\sum_{k}\varepsilon_{k}c_{ok}^{\dagger}c_{ok}-V_{BT}\sum_{kp}c_{ok}^{\dagger}c_{op}\label{eq:odd}
\end{equation}
is the decoupled one. However, they are connected to each other by
the tunneling Hamiltonian $\mathcal{\tilde{H}}_{\text{{tun}}}=-\Delta\mu\sum_{k}(c_{ek}^{\dagger}c_{ok}+c_{ok}^{\dagger}c_{ek}).$

As in the zero-bias regime $\Delta\mu\rightarrow0$, due to $\varphi\rightarrow0$,
$\mathcal{\tilde{H}}_{\text{{tun}}}$ is a perturbative term and the
linear response theory ensures that
\begin{equation}
\mathcal{T}\left(\varepsilon\right)=(2\pi V_{BT})^{2}\tilde{\rho}_{e}(\varepsilon)\tilde{\rho}_{o}(\varepsilon),\label{eq:transmit-1}
\end{equation}
where $\tilde{\rho}_{e}(\varepsilon)=-\frac{1}{\pi}{\tt Im}(\tilde{\mathcal{G}}_{\psi_{e}\psi_{e}})$
is the local density of states (LDOS) for the Hamiltonian of Eq.\,(\ref{eq:even})
and
\begin{align}
\mathcal{G}_{\Psi_{e}\Psi_{e}} & =-\frac{i}{\hbar}\theta\left(t\right){\tt Tr}\{\varrho_{\text{e}}[\Psi_{e}\left(t\right),\Psi_{e}^{\dagger}\left(0\right)]_{+}\}\label{eq:PSI_R-2}
\end{align}
gives the retarded Green's function in the time domain $t$, where
$\theta(t)$ is the Heaviside step function, $\varrho_{\text{e}}$
is the density-matrix for Eq.\,($\ref{eq:even}$), $\Psi_{e}=f_{e}+(\pi\Delta\rho_{0})^{1/2}q\sum_{j}d_{j}$
is a field operator, with $f_{e}=\sum_{p}c_{ep},$ the Anderson parameter
$\Delta=2\pi V^{2}\rho_{0}$ and $q=\left(\pi\rho_{0}\Delta\right)^{-1/2}\left(\frac{\sqrt{2}V}{2V_{BT}}\right).$

To calculate Eq.\,(\ref{eq:PSI_R-2}) in the energy domain $\varepsilon$,
we should employ the equation-of-motion (EOM) method \cite{EOM,book} summarized as
follows
\begin{equation}
(\varepsilon+i0^{+})\tilde{\mathcal{G}}_{\mathcal{AB}}=[\mathcal{A},\mathcal{B^{\dagger}}]_{+}+\tilde{\mathcal{G}}_{\left[\mathcal{A},\mathcal{\mathcal{H}}_{i}\right]\mathcal{B}}\label{eq:EOM}
\end{equation}
for the retarded Green's function $\tilde{\mathcal{G}}_{\mathcal{AB}},$
with $\mathcal{A}$ and $\mathcal{B}$ as fermionic operators belonging
to the Hamiltonian $\mathcal{\mathcal{H}}_{i}.$ By considering $\mathcal{A=B}=\Psi{}_{e}$
and $\mathcal{H}_{i}=\mathcal{H}_{e}$, we find
\begin{align}
\tilde{\mathcal{G}}_{\Psi_{e}\Psi_{e}} & =\tilde{\mathcal{G}}_{f_{e}f_{e}}+(\pi\rho_{0}\Delta)q^{2}\sum_{jl}\tilde{\mathcal{G}}_{d_{j}d_{l}}+2(\pi\rho_{0}\Delta)^{1/2}q\nonumber \\
 & \times\sum_{j}\tilde{\mathcal{G}}_{d_{j}f_{e}}.\label{eq:25b}
\end{align}
From Eqs.\,(\ref{eq:even}),
(\ref{eq:EOM}) with $\mathcal{A=B}=f_{e}$ and (\ref{eq:25b}), we obtain
\begin{align}
\tilde{\mathcal{G}}_{f_{e}f{}_{e}} & =\frac{\pi\rho_{0}(\bar{q}-i)}{1-\sqrt{x}(\bar{q}-i)}+\pi\rho_{0}\Delta\left[\frac{(\bar{q}-i)}{1-\sqrt{x}(\bar{q}-i)}\right]^{2}\nonumber \\
 & \times\sum_{jl}\tilde{\mathcal{G}}_{d_{j}d_{l}}\left(\varepsilon\right)\label{eq:g_ff}
\end{align}
and the mixed Green's function
\begin{equation}
\tilde{\mathcal{G}}_{d_{j}f_{e}}=\sqrt{\pi\Delta\rho_{0}}\frac{(\bar{q}-i)}{1-\sqrt{x}(\bar{q}-i)}\sum_{l}\tilde{\mathcal{G}}_{d_{j}d_{l}},\label{eq:g_df}
\end{equation}
determined from Eq.\,(\ref{eq:EOM}) by considering $\mathcal{A}=d_{j}$,
$\mathcal{B}=f{}_{e}$ and $\mathcal{H}_{i}=\mathcal{H}_{e}$, with
the parameter $x=(\pi\rho_{0}V_{BT})^{2}$ and $\bar{q}=\frac{1}{\pi\rho_{0}}\sum_{k}\frac{1}{\varepsilon-\varepsilon_{k}}.$

Additionally, for the Hamiltonian of Eq.\,(\ref{eq:odd}) we have the
LDOS $\tilde{\rho}_{o}(\varepsilon)=-\frac{1}{\pi}{\tt Im}(\tilde{\mathcal{G}}_{f_{o}f_{o}}),$
with
\begin{align}
\mathcal{G}_{f_{o}f_{o}} & =-\frac{i}{\hbar}\theta\left(t\right){\tt Tr}\{\varrho_{\text{o}}[f_{o}\left(t\right),f_{o}^{\dagger}\left(0\right)]_{+}\}\label{eq:PSI_R-2-1}
\end{align}
and $f_{o}=\sum_{\tilde{q}}c_{o\tilde{q}}.$ We notice that $\tilde{\mathcal{G}}_{f_{o}f_{o}}$ is decoupled from
the QDs. Thereby, from Eqs.\,(\ref{eq:odd}) and (\ref{eq:PSI_R-2-1}),
we take $\mathcal{A}=\mathcal{B}=f{}_{o}$ and $\mathcal{H}_{i}=\mathcal{H}_{o}$
in Eq.\,(\ref{eq:EOM}) and we obtain
\begin{equation}
\tilde{\mathcal{G}}_{f_{o}f{}_{o}}=\frac{\pi\rho_{0}(\bar{q}-i)}{1+\sqrt{x}(\bar{q}-i)}.\label{eq:rho_ff}
\end{equation}
Thus the substitution of Eqs.\,(\ref{eq:25b}), (\ref{eq:g_df}),
and (\ref{eq:rho_ff}) in Eq.\,(\ref{eq:transmit-1}), leads to
\begin{align}
\mathcal{T}\left(\varepsilon\right) & =\mathcal{T}_{b}+\sqrt{\mathcal{T}_{b}\mathcal{R}_{b}}\tilde{\Delta}\sum_{j\tilde{j}}{\tt Re}\{\tilde{\mathcal{G}}_{d_{j}d_{\tilde{j}}}\left(\varepsilon\right)\}\nonumber \\
 & -(1-2\mathcal{T}_{b})\frac{\tilde{\Delta}}{2}\sum_{j\tilde{j}}{\tt Im}\{\tilde{\mathcal{G}}_{d_{j}d_{\tilde{j}}}\left(\varepsilon\right)\},\label{eq:trans2}
\end{align}
where $\tilde{\Delta}=\frac{\Delta}{1+x}$ is an effective dot-lead
coupling, $\mathcal{T}_{b}=\frac{4x}{\left(1+x\right)^{2}}$ represents
the background transmittance and $\mathcal{R}_{b}=1-\mathcal{T}_{b}=\frac{\left(1-x\right)^{2}}{\left(1+x\right)^{2}}$
is the corresponding reflectance, both in the absence of the QDs and MBSs.

We emphasize that Eq.\,(\ref{eq:trans2}) is the generalization
of Eq.\,(2) found in Ref.\,{[}\onlinecite{key-26}{]} for an Aharonov-Bohm-like device of a single QD, but without an applied magnetic field.
Yet, in Ref.\,{[}\onlinecite{key-26}{]} the authors introduce $q_{b}=\sqrt{\frac{\mathcal{R}_{b}}{\mathcal{T}_{b}}}=\frac{\left(1-x\right)}{2\sqrt{x}}$
as the Fano parameter. Here we focus on two cases: the case
$q_{b}\rightarrow\infty$, which we call by regime of weak coupling
lead-lead due to the background transmittance $\mathcal{T}_{b}=0$,
and the strong coupling limit $q_{b}=0$ characterized by $\mathcal{T}_{b}=1$.

We calculate the Green's functions $\tilde{\mathcal{G}}_{d_{j}d_{l}}$
within the wide-band limit. To obtain them, we first express the
Majorana operators $\Psi_{1}$ and $\Psi_{2}$ in terms of a nonlocal
Dirac fermion state $\eta$ as follows: $\Psi_{1}=\frac{1}{\sqrt{2}}(\eta^{\dagger}+\eta)$
and $\Psi_{2}=i\frac{1}{\sqrt{2}}(\eta^{\dagger}-\eta),$ with $\eta\neq\eta^{\dagger}$
and $\left\{ \eta,\eta^{\dagger}\right\} =1$. We verify that Eq.\,(\ref{eq:MBSs}) transforms into
\begin{eqnarray}
\mathcal{H}_{\text{{MBSs}}} & = & \varepsilon_{M}(\eta^{\dagger}\eta-\frac{1}{2})+\frac{\lambda}{\sqrt{2}}(d_{1}\eta^{\dagger}+\eta d_{1}^{\dagger})\nonumber \\
 & + & \frac{\lambda}{\sqrt{2}}(d_{1}\eta-d_{1}^{\dagger}\eta^{\dagger}).\label{eq:new_MBSs}
\end{eqnarray}

By applying the EOM on
\begin{align}
\mathcal{G}_{d_{j}d_{l}} & =-\frac{i}{\hbar}\theta\left(t\right){\tt Tr}\{\varrho_{\text{{e}}}[d_{j}\left(t\right),d_{l}^{\dagger}\left(0\right)]_{+}\},\label{eq:djdl}
\end{align}
and changing to the energy domain $\varepsilon$, we obtain the following
relation:
\begin{align}
(\varepsilon-\varepsilon_{j}-\Sigma-\delta_{j1}\Sigma_{\text{{MBS1}}})\tilde{\mathcal{G}}_{d_{j}d_{l}} & =\delta_{jl}+\Sigma\sum_{\tilde{l}\neq j}\tilde{\mathcal{G}}_{d_{\tilde{l}}d_{l}},\label{eq:GjjA}
\end{align}
with $\Sigma=-\frac{(\sqrt{x}+i)}{1+x}\Delta$ and $\Sigma_{\text{{MBS1}}}=\lambda^{2}K(1+\lambda^{2}\tilde{K})$
as the self-energy due to the MBS 1 coupled to the QD 1, $K=\frac{1}{2}\left(\frac{1}{\varepsilon-\varepsilon_{M}+i0^{+}}+\frac{1}{\varepsilon+\varepsilon_{M}+i0^{+}}\right)$
and $\tilde{K}=\frac{K}{\varepsilon+\varepsilon_{1}+\bar{\Sigma}-\lambda^{2}K}$
have the same forms as found in Ref.\,{[}\onlinecite{key-25}{]},
where $\bar{\Sigma}$ is the complex conjugate of $\Sigma$. Thus
the solution of Eq.\,(\ref{eq:GjjA}) provides
\begin{align}
\tilde{\mathcal{G}}_{d_{1}d_{1}} & =\frac{1}{\varepsilon-\varepsilon_{1}-\Sigma-\Sigma_{\text{{MBS1}}}-\mathcal{C}_{2}}\nonumber \\
\label{eq:d1d1}
\end{align}
as the Green's function of the QD 1, with $\mathcal{C}_{j}=\frac{\Sigma^{2}}{\varepsilon-\varepsilon_{j}-\Sigma}$
as the self-energy due to the presence of the \textit{$j^{th}$} QD.

For $\mathcal{C}_{2}=0$, we highlight that Eq.\,(\ref{eq:d1d1}) is
reduced to the Green's function of the single QD system found in Ref.\,{[}\onlinecite{key-25}{]}. In the case of the QD 2, we have
\begin{align}
\tilde{\mathcal{G}}_{d_{2}d_{2}} & =\frac{1-\tilde{\mathcal{G}}_{d_{1}d_{1}}^{0}\Sigma_{\text{{MBS1}}}}{\varepsilon-\varepsilon_{2}-\Sigma-\dfrac{\tilde{\mathcal{G}}_{d_{1}d_{1}}^{0}}{\tilde{\mathcal{G}}_{d_{2}d_{2}}^{0}}\Sigma_{\text{{MBS1}}}-\mathcal{C}_{1}},\label{eq:d2d2}
\end{align}
where $\tilde{\mathcal{G}}_{d_{1}d_{1}}^{0}=1/(\varepsilon-\varepsilon_{1}-\Sigma)$
and $\tilde{\mathcal{G}}_{d_{2}d_{2}}^{0}=1/(\varepsilon-\varepsilon_{2}-\Sigma)$
represent the corresponding Green's functions for the single QD system
without Majoranas. The mixed Green's functions are
\begin{equation}
\tilde{\mathcal{G}}_{d_{2}d_{1}}=\frac{\Sigma}{\varepsilon-\varepsilon_{2}-\Sigma}\tilde{\mathcal{G}}_{d_{1}d_{1}}\label{eq:d2d1}
\end{equation}
and
\begin{align}
\tilde{\mathcal{G}}_{d_{1}d_{2}} & =\frac{\Sigma}{\varepsilon-\varepsilon_{1}-\Sigma-\Sigma_{\text{{MBS1}}}}\tilde{\mathcal{G}}_{d_{2}d_{2}}.\label{eq:d1d2}
\end{align}

By inspecting Eqs.\,(\ref{eq:d1d1}), (\ref{eq:d2d2}), (\ref{eq:d2d1})
and (\ref{eq:d1d2}), we verify that the functions $\tilde{\mathcal{G}}_{d_{1}d_{1}}$
and $\tilde{\mathcal{G}}_{d_{1}d_{2}}$ can be found by the exchange
of the indexes $1\leftrightarrow2$ in $\tilde{\mathcal{G}}_{d_{2}d_{2}}$
and $\tilde{\mathcal{G}}_{d_{2}d_{1}}$, respectively, just in the
case of $\Sigma_{\text{{MBS1}}}=0.$ Nevertheless, in the opposite
situation, namely for $\Sigma_{\text{{MBS1}}}\neq0$, this symmetry is not
verified, which corresponds to the Kitaev wire coupled to the interferometer
as outlined in Fig.\,\ref{Fig1}.

Following Ref.\,\cite{key-25}, we compare the aforementioned symmetry breaking with the one that arises from a Regular Fermionic (RF) zero mode attached to the interferometer instead of the MBS 1. In such a case, one can show that $\Sigma_{\text{{MBS1}}}$ is replaced by $\Sigma_{\text{{RF}}}=\frac{\lambda^{2}}{\varepsilon+i0^{+}}$ in the Green's functions above. Thus the comparison proposed will allow us to isolate exclusively Majorana signatures from those due to a standard QD side-coupled to the interferometer.

It is worth mentioning that the Green's functions for the QDs derived in this work are exact as expected since the Hamiltonian of Eq.\,(\ref{eq:TIAM}) is quadratic. From the experimental point of view this feature can be feasible by keeping the two QDs far apart in such a way that the interdots Coulomb repulsion $U$ becomes negligible to ensure the applicability of the present approach, otherwise the term $Ud_{1}^{\dagger}d_{1}d_{2}^{\dagger}d_{2}$ should be added to Eq.\,(\ref{eq:TIAM}). In this situation, an interacting self-energy $\Sigma_{\text{{U}}}$ enters in the Green's functions and leads to a dephasing rate $-\text{Im}(\Sigma_{\text{{U}}})/{\hbar}$ that competes with $-\text{Im}(\Sigma)/{\hbar},$ $-\text{Im}(\Sigma_{\text{{MBS1}}})/{\hbar}$ and $-\text{Im}(\mathcal{C}_{j})/{\hbar}.$ Particularly if $U$ is relevant in the regime of extremely low temperatures, it will be possible to induce a Kondo effect \cite{key-27} and a ZBA  determined by the interplay between such a phenomenon and the Majorana zero mode should appear. To avoid the emergence of the Kondo resonance in the ZBA  of the transmittance we thereby assume two QDs far apart to fulfill the assumption $U=0$.

\section{Results and Discussion}
\label{sec3}

Below we investigate the
antisymmetric feature of the Green's functions by employing the expression for the transmittance (Eq.\,(\ref{eq:trans2}))
with $\lambda=4\Delta$ in the case of the Kitaev wire coupled
to the interferometer. By varying $\varepsilon$ in Eq.\,(\ref{eq:trans2}),
the transmittance $\mathcal{T}$ becomes a function of the Fermi energy
$\varepsilon=\mu_{T}=\mu_{B}$ of the leads \cite{JAP1,JAP2}. According
to Eq.\,(\ref{eq:_10b}), this transmittance can be obtained experimentally
via the conductance $G$ in units of $G_{0}=e^{2}/h$ for temperatures
$T\rightarrow0$ just by attaching gate voltages to the metallic leads
in order to tune the Fermi level. Additionally, we estimate the symmetric
detuning $\Delta\varepsilon,$ the Fermi energy $\varepsilon$ and
$\varepsilon_{M}$ in units of the Anderson parameter $\Delta.$

\begin{figure}
\includegraphics[width=0.46\textwidth,height=0.2\textheight]{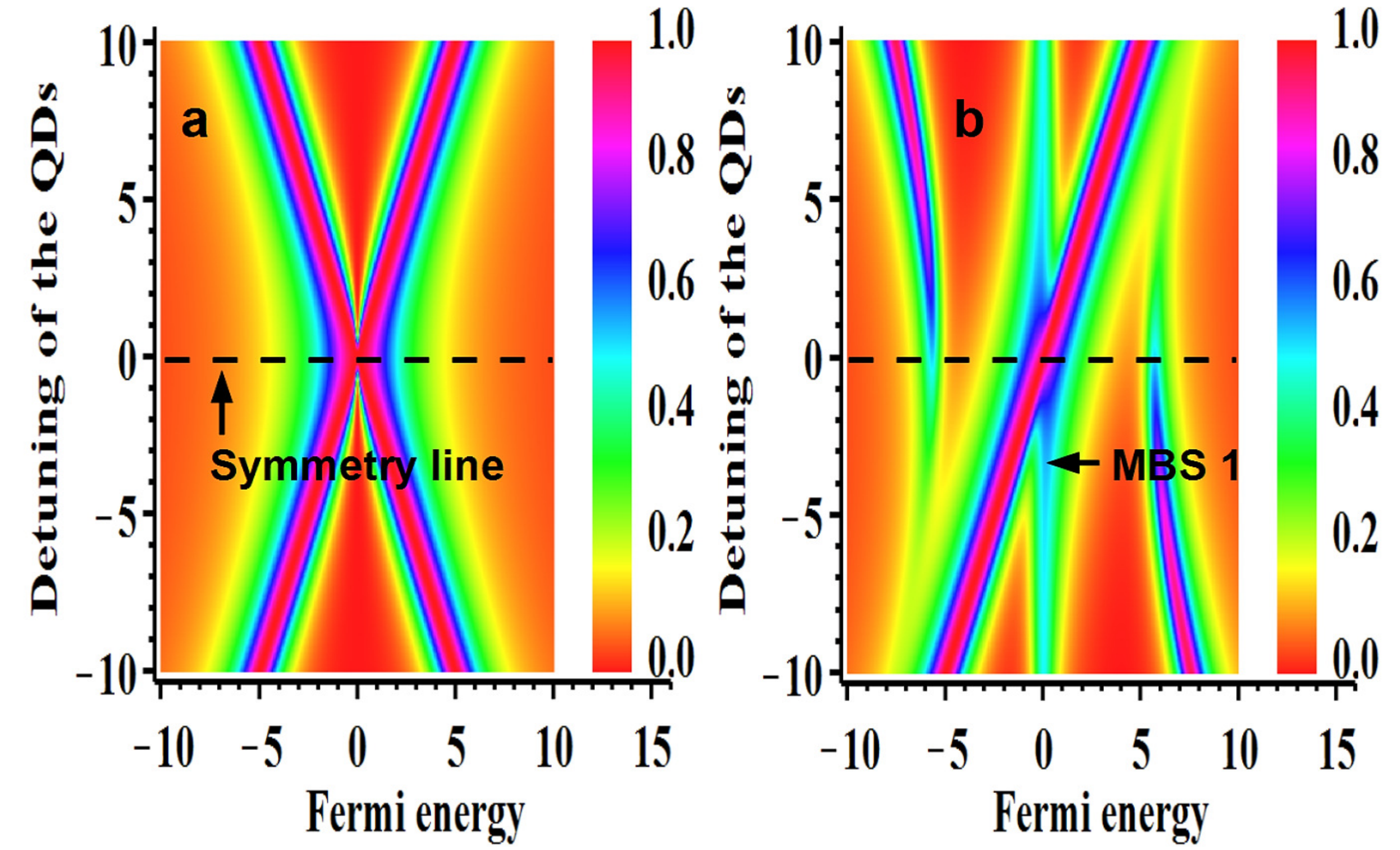}
\caption{\label{Fig2}(Color online) Density plot of the transmittance (Eq.\,(\ref{eq:trans2})) as a function of the symmetric detuning $\Delta\varepsilon$
for the QDs and the Fermi energy $\varepsilon$ in units of $\Delta$
within the Fano regime $x=0$ $(q_{b}\rightarrow\infty)$: (a) The
Aharonov-Bohm-like interferometer is decoupled from the Kitaev wire
leading to a symmetric Fano effect. (b) The interferometer is hybridized
with a semi-infinite Kitaev wire via the QD 1: the Fano effect is
antisymmetric. The central structure at $\varepsilon=0$ gives the
ZBA due to the MBS 1. In both panels the orange color denote a perfect
insulating behavior.}
\end{figure}

In Fig.\,\ref{Fig2}, the density plot of the transmittance is presented
as a function of the symmetric detuning $\Delta\varepsilon$ and
the Fermi level $\varepsilon$ for the Fano regime $x=0$ $(q_{b}\rightarrow\infty)$.
The panel (a) exhibits the situation in which the Kitaev wire is decoupled
from the interferometer, thus leading to the symmetric Fano interference
arising from symmetric Green's functions by the permutation of the
indexes in the parameters of the QDs. As a result, the density plot
is characterized by specular regions with respect to the dotted-black
line labeled as symmetry line.

On the other hand, for a semi-infinite
Kitaev wire $(\varepsilon_{M}=10^{-7})$ attached to the interferometer,
the specular feature is not observed as the aftermath of the Green's
functions $\tilde{\mathcal{G}}_{d_{j}d_{j}}$ and $\tilde{\mathcal{G}}_{d_{j}d_{l}}$
for the QDs, which cannot be determined by the exchange of the indexes
$j\leftrightarrow l$ in $\tilde{\mathcal{G}}_{d_{l}d_{l}}$ and $\tilde{\mathcal{G}}_{d_{l}d_{j}}$,
respectively, with $l,j=1...2.$ Such a hallmark is shown in the panel
(b) of Fig.\,\ref{Fig2}, where we clearly visualize the distortion of the pattern found
in panel (a): the antisymmetric Fano effect for the limit $x=0$ $(q_{b}\rightarrow\infty)$
assisted by the isolated MBS 1 can be recognized by the current antisymmetric
pattern of (b). In the latter, the central region is the ZBA due to
the MBS 1, which occurs for $\varepsilon=0$ and any value of $\Delta\varepsilon$.

\begin{figure}
\includegraphics[width=0.46\textwidth,height=0.2\textheight]{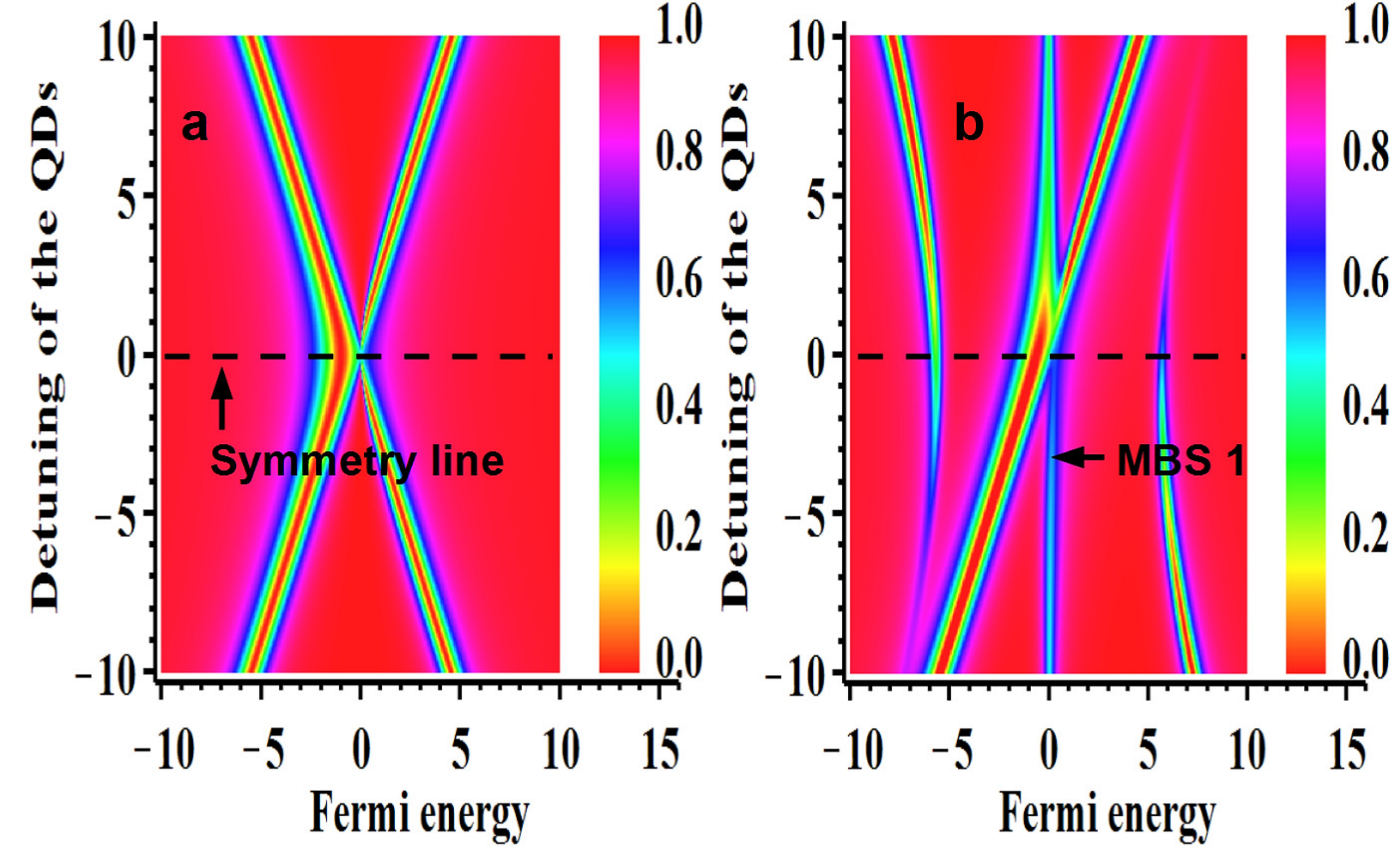}
\caption{\label{Fig3}(Color online) Density plot of the transmittance (Eq.\,(\ref{eq:trans2})) as a function of the symmetric detuning $\Delta\varepsilon$
for the QDs and the Fermi energy $\varepsilon$ in units of $\Delta$
for the opposite Fano regime $x=1$ $(q_{b}=0)$:
the central structure at $\varepsilon=0$ gives the reversed ZBA due
to the MBS 1. In panels (a) and (b) the red regions denote a perfect
conducting behavior, which is the reversed of that found in Fig.\,\ref{Fig2}.}
\end{figure}

\begin{figure}
\includegraphics[width=0.46\textwidth,height=0.2\textheight]{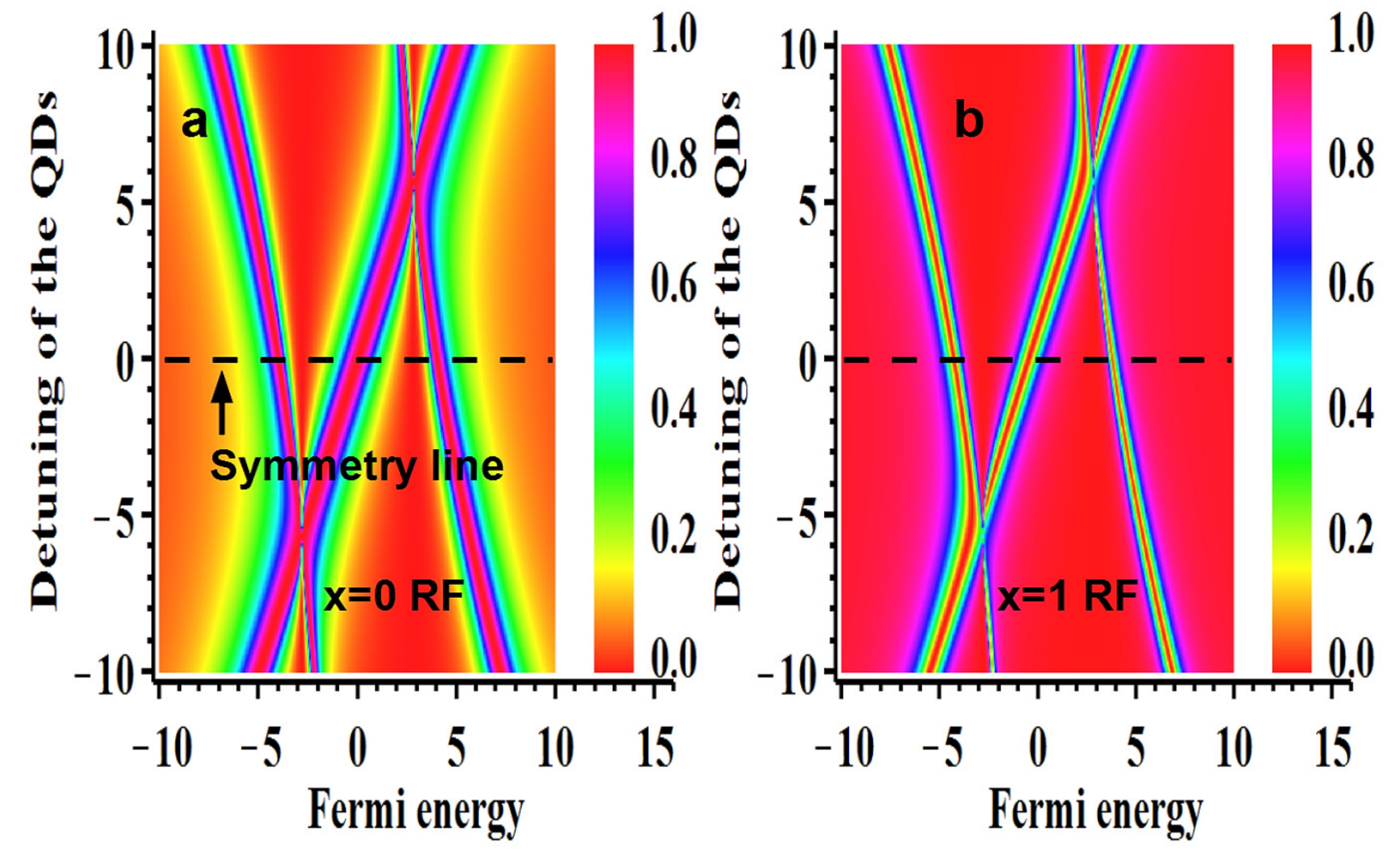}
\caption{\label{Fig4}(Color online) Density plot of the transmittance (Eq.\,(\ref{eq:trans2})) as a function of the symmetric detuning $\Delta\varepsilon$
for the QDs and the Fermi energy $\varepsilon$ in units of $\Delta$
for a RF zero mode coupled to the interferometer in opposite Fano regimes: (a) $x=0$ $(q_{b}\rightarrow\infty)$ and (b) $x=1$ $(q_{b}=0).$}
\end{figure}

Fig.\,\ref{Fig3} exhibits the opposite Fano regime, which is determined
by $x=1$ $(q_{b}=0).$ In this limit, the tunneling of electrons through the leads is dominant with respect to that via the QDs. In panel (a)
of Fig.\,\ref{Fig3} for the Kitaev wire removed, we observe the specular
feature as the corresponding found in Fig.\,\ref{Fig2}(a), but with
the regime of interference reversed: note that the orange color regions become
replaced by the red type. This swap indicates that a perfect insulating
behavior changes to a conducting one as pointed out in Fig.\,\ref{Fig3}(a).

In the presence of a semi-infinite Kitaev wire, the reversed pattern of
Fig.\,\ref{Fig2}(b) is given by Fig.\,\ref{Fig3}(b), where the antisymmetric
Fano effect manifests itself via the antisymmetric upper and lower
parts of this figure. As a result, the difference between panels (a)
and (b) of Figs.\,\ref{Fig2} and \ref{Fig3} reveals that the emergence
of new insulating and conduction regions with respect to those found
in panels (a) of both figures can be considered as fingerprints of
an isolated MBS. Additionally, the ZBA due to the MBS 1 is also reversed.
It is worth mentioning that the remaining diagonals in Figs.\,\ref{Fig2}(b) and \,\ref{Fig3}(b) are due to the QD 2, which even decoupled from the MBS 1 is still sensitive to it as Eq.\,(\ref{eq:d2d2}) ensures via the self-energy $\Sigma_{\text{{MBS1}}}=\lambda^{2}K(1+\lambda^{2}\tilde{K}).$ As we can notice, such a dependence yields a slight change in the slopes of these diagonals with respect to those observed in Figs.\,\ref{Fig2}(a) and \,\ref{Fig3}(a) obtained with $\Sigma_{\text{{MBS1}}}=0.$

\begin{figure}
\includegraphics[width=0.46\textwidth,height=0.2\textheight]{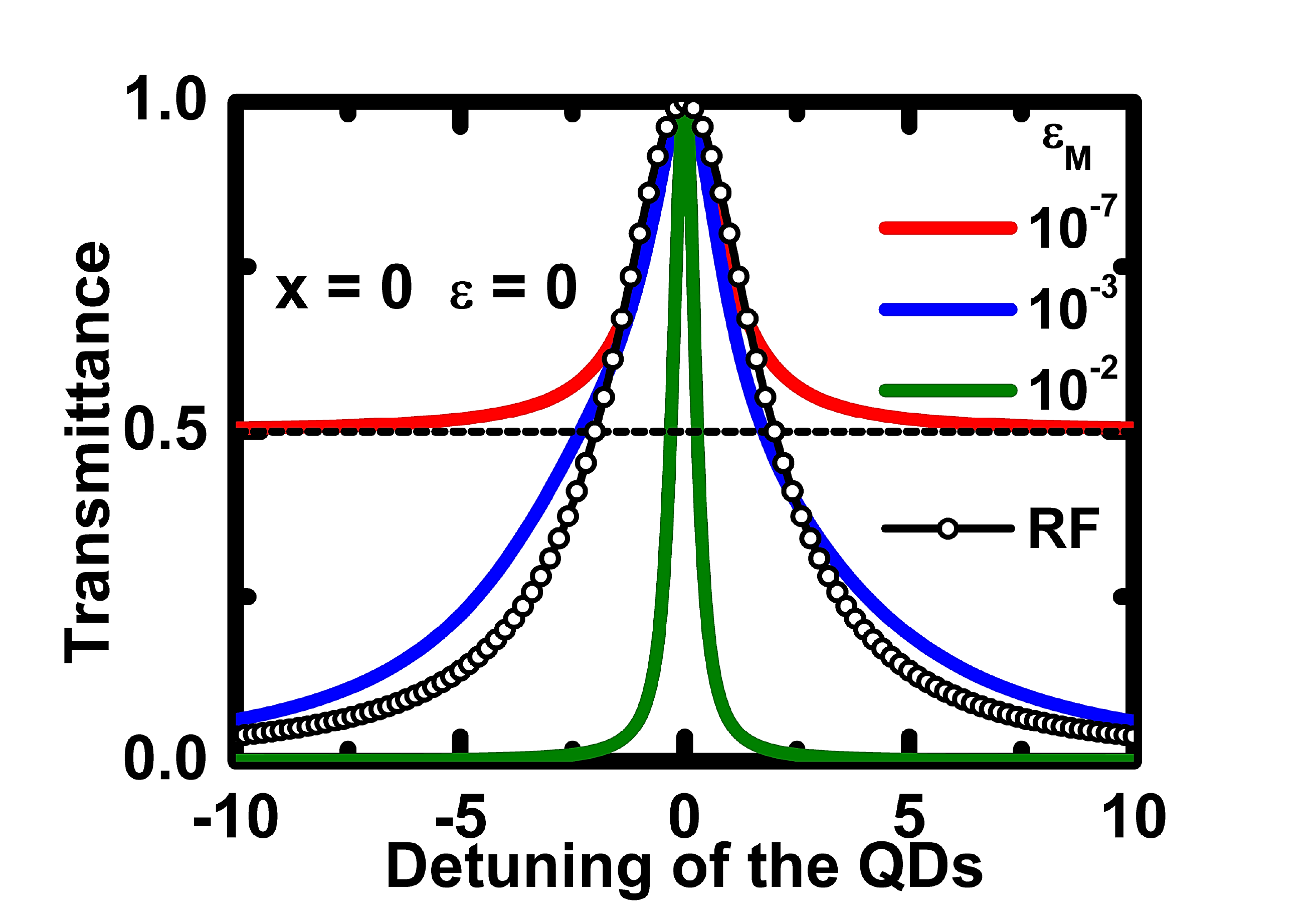}
\caption{\label{Fig5}(Color online) Transmittance of Eq.\,(\ref{eq:trans2})
as a function of the symmetric detuning $\Delta\varepsilon$ with
Fermi energy $\varepsilon=0$ in units of $\Delta$ within the symmetric
Fano regime $x=0$ for different lengths $L$ of the Kitaev wire given by the parameter
$\varepsilon_{M}\sim e^{-L/\xi}$ in Eq.\,(\ref{eq:MBSs}) as well as for the case of a RF zero mode. }
\end{figure}

\begin{figure}[!]
\centerline{\resizebox{3.3in}{!}{\includegraphics[clip,width=0.14\textwidth,height=0.06\textheight]{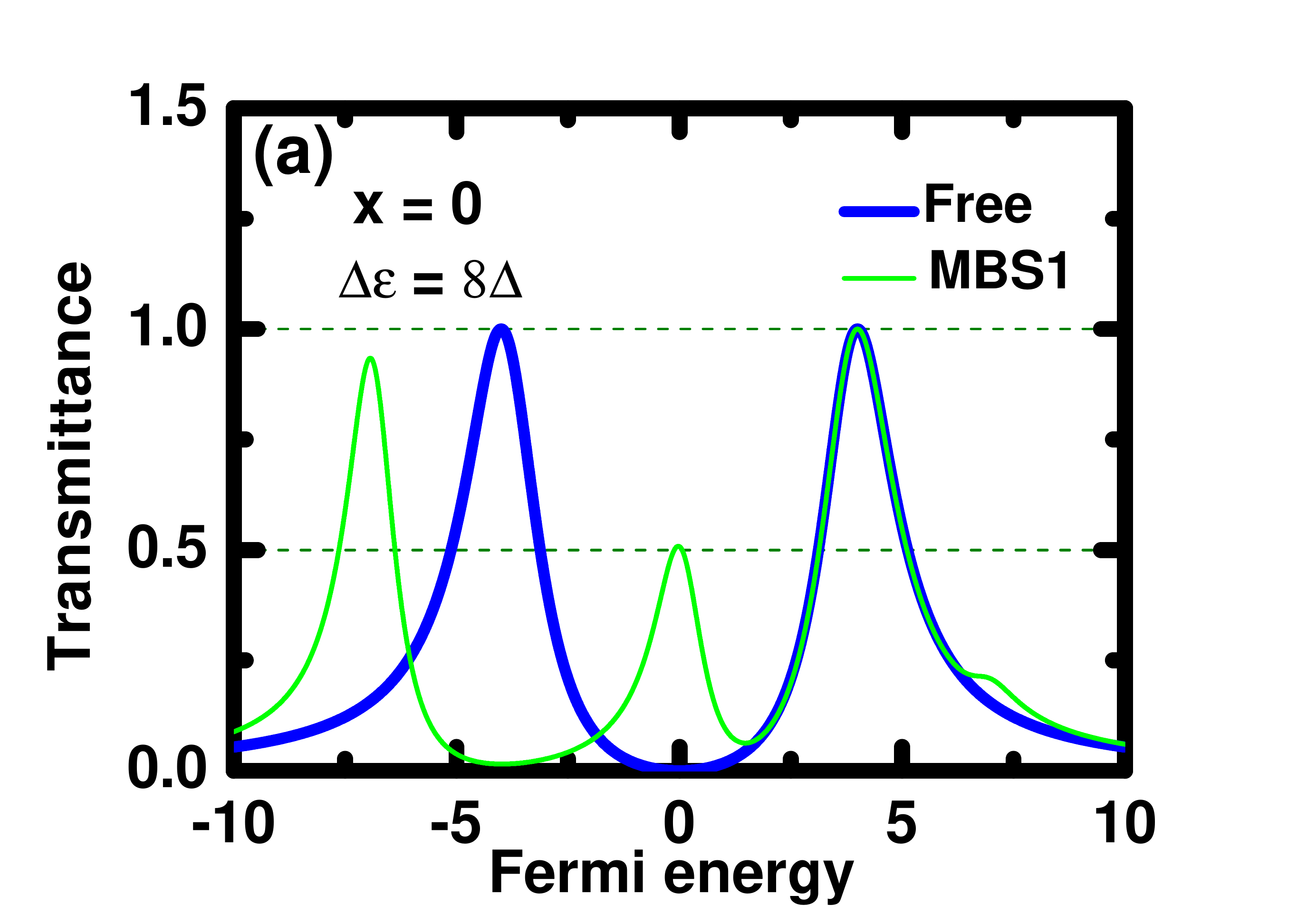}}}\centerline{\resizebox{3.3in}{!}{\includegraphics[clip,width=0.14\textwidth,height=0.06\textheight]{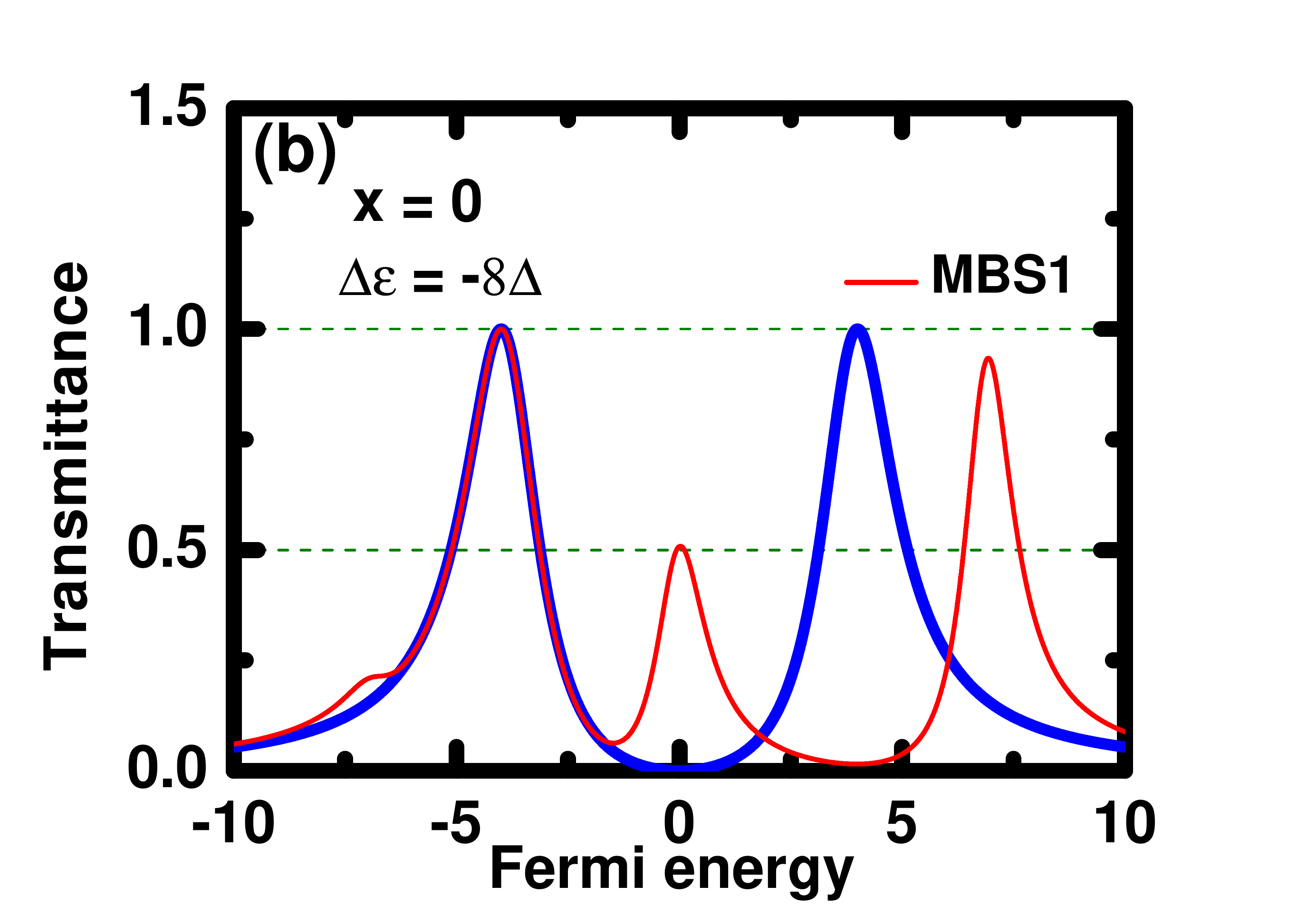}}}
\caption{(Color online) Transmittance determined by Eq.\,(\ref{eq:trans2}) as a function of the Fermi energy $\varepsilon$ in units of $\Delta$ in the Fano regime $x=0$ $(q_{b}\rightarrow\infty)$ for the double QD setup of Fig.\,\ref{Fig1} with $\varepsilon_{M}=10^{-7}$. In panel (a) we use $\Delta\varepsilon=8\Delta$: the solid-green lineshape is for the situation of the MBS 1 present, while the blue curve corresponds to the free interferometer in which the Kitaev wire is absent. Panel (b): for the reversed detuning $\Delta\varepsilon=-8\Delta,$ only the pronounced satellite peaks of the QDs in the free case (solid-blue lineshape) and the ZBA in the solid-red lineshape remain placed at the same positions with respect to those found in panel (a). In both panels a Majorana peak (ZBA) rises characterized by an amplitude nearby $1/2$. Such a central structure is robust against the gates swap in contrast to the pronounced satellite peaks of the QDs when the Kitaev wire is side-coupled to one of these dots.}
\label{Fig6}
\end{figure}

Fig.\,\ref{Fig4} shows the analysis of the transmittance for the Fano limits $x=0$ and $x=1$ in the case of a RF zero mode coupled to the interferometer via QD 1. As we can verify, the symmetry breaking is distinct with respect to that observed in the situation of a Kitaev wire present: the patterns of Figs.\,\ref{Fig2}(b) and \ref{Fig3}(b) do not match with those found in Figs.\,\ref{Fig4}(a) and (b), respectively. Consequently, this difference ensures that the patterns of Figs.\,\ref{Fig2}(b) and \ref{Fig3}(b) constitute robust signatures due to an isolated MBS. In which concerns on the remaining diagonals of the QD 2 for the RF zero mode, the change in the slopes are weaker than for the MBS 1, since in the RF situation the self-energy $\Sigma_{\text{{RF}}}$ is just proportional to $\lambda^2.$

In Fig.\,\ref{Fig5}, we present the transmittance through the interferometer
for the leads in resonance with the MBS 1 for the case
of weak coupling lead-lead characterized by $x=0$ $(q_{b}\rightarrow\infty)$.
This regime corresponds to the situation whereby electrons travel
exclusively via the QDs, which are also in resonance with the MBS
1. Two distinct wire lengths are analyzed: (i) the semi-infinite Kitaev
wire and (ii), the case of a finite system. For the setup (i) given
by $\varepsilon_{M}=10^{-7},$ it is possible to notice in the red
lineshape that the transmittance approaches half for $|\Delta\varepsilon|\rightarrow\infty,$
while it evolves towards the unity in the limit $|\Delta\varepsilon|\rightarrow0$.

The unitary value of the transmittance occurs for the QDs in resonance
with the MBS 1 $(\Delta\varepsilon=0)$, thus leading to the maximum
transmittance, since the two QDs and the MBS 1 allow a perfect resonant
tunneling through the Fermi level of the leads. Away from $\Delta\varepsilon=0,$
the transmittance becomes half the unity as a result of the isolated
MBS 1, which is a half-electron state (see the semi-sphere in the
left side of the Kitaev wire of Fig.\,\ref{Fig1}) and the only piece
of the system in resonance with the Fermi level of the leads. For
$0<|\Delta\varepsilon|<\infty,$ the resonant tunneling due to the
QDs is not perfect and the transmittance stays within the range $1/2<\mathcal{T}<1.$
For the setup (ii), an unitary fluctuation in the transmittance as
a function of $\Delta\varepsilon$ emerges in opposite to the value
of half observed in the system (i): see the curves given by the blue
and green colors, respectively for $\varepsilon_{M}=10^{-3}$ and
$\varepsilon_{M}=10^{-2}.$ In both profiles, we verify that the transmittance
reaches the unity for $\Delta\varepsilon=0,$ but it vanishes as we
increase $|\Delta\varepsilon|.$

The variation of the maximum value of the transmittance has a simple physical
origin: it arises from the presence of the MBS 2 (see the semi-sphere
in the right side of the Kitaev wire of Fig.\,\ref{Fig1}), which combines
with the MBS 1 in order to form the regular and delocalized fermionic
state $\eta$ as Eq.\,(\ref{eq:new_MBSs}) shows. Such a regular fermion
adds an extra factor of half to the fluctuation of $\mathcal{T}$
and provides the unitary suppression of the transmittance for $|\Delta\varepsilon|\rightarrow\infty.$
Additionally, the bigger $\varepsilon_{M}$ the shorter the length of
the Kitaev wire according to the term $\varepsilon_{M}\sim e^{-L/\xi}$
in Eq.\,(\ref{eq:MBSs}) and sharper the widths of the profiles of
the transmittance as a function of $\Delta\varepsilon$ (notice that
the green line is sharper than the corresponding blue one).

In situations (i) and (ii), the Fano effect is symmetric since the
left and right sides of Fig.\,\ref{Fig5} are specular. We also show in Fig.\,\ref{Fig5} the interferometer hybridized with a RF zero mode instead of a MBS 1. Such a curve is given by the black-circle lineshape, in which we can notice that the fluctuation of the transmittance is still one as expected from the regular nature of the electronic state of the QD. However, the width of the resonance differs from those found in the finite wires analyzed as the aftermath of the functional form of the self-energy $\Sigma_{\text{{RF}}}$, which is size independent in opposite to $\Sigma_{\text{{MBS1}}}.$

In Fig.\,\ref{Fig6} we analyze the profiles of the transmittance as a function of the Fermi energy for fixed detunings $\Delta\varepsilon$ within the Fano limit $x=0$ $(q_{b}\rightarrow\infty)$ in the situations of the semi-infinite Kitaev wire $(\varepsilon_{M}=10^{-7})$ present and absent. To that end we account two distinct scenarios: (a) $\Delta\varepsilon>0$ and (b) $\Delta\varepsilon<0.$ The features of the former are printed in panel (a) where we can clearly visualize a couple of pronounced satellite peaks both in the solid-blue and green lineshapes, which are due to the imposed detuning $\Delta\varepsilon=8\Delta$ for the QDs. Note that in the MBS $1$ case (solid-green lineshape) the detuning is bigger than that observed in the free case (solid-blue lineshape of panel (a)) as expected due to the renormalization of the level $\varepsilon_{1}=-4\Delta$ of the QD $1$ hybridized with the Kitaev wire, since the QD $2$ with $\varepsilon_{2}=4\Delta$ is decoupled from it as outlined in Fig.\,\ref{Fig1}. For panel (b) in which $\Delta\varepsilon=-8\Delta$, the positions of the pronounced satellite peaks in the transmittance for the free interferometer (blue curve) as well as that of the ZBA in the solid-red lineshape (MBS $1$ case) persist as those observed in panel (a). On the other hand, the pronounced satellite peaks in presence of the MBS $1$ (red curve) do not as a result of the symmetry-breaking feature previously discussed for the system Green's functions. We stress that the ZBA is then immune against the swapping of gates thus revealing the robustness of the MBS 1, being characterized by a transmittance amplitude close to $1/2$.

\begin{figure}[!]
\centerline{\resizebox{3.3in}{!}{\includegraphics[clip,width=0.14\textwidth,height=0.06\textheight]{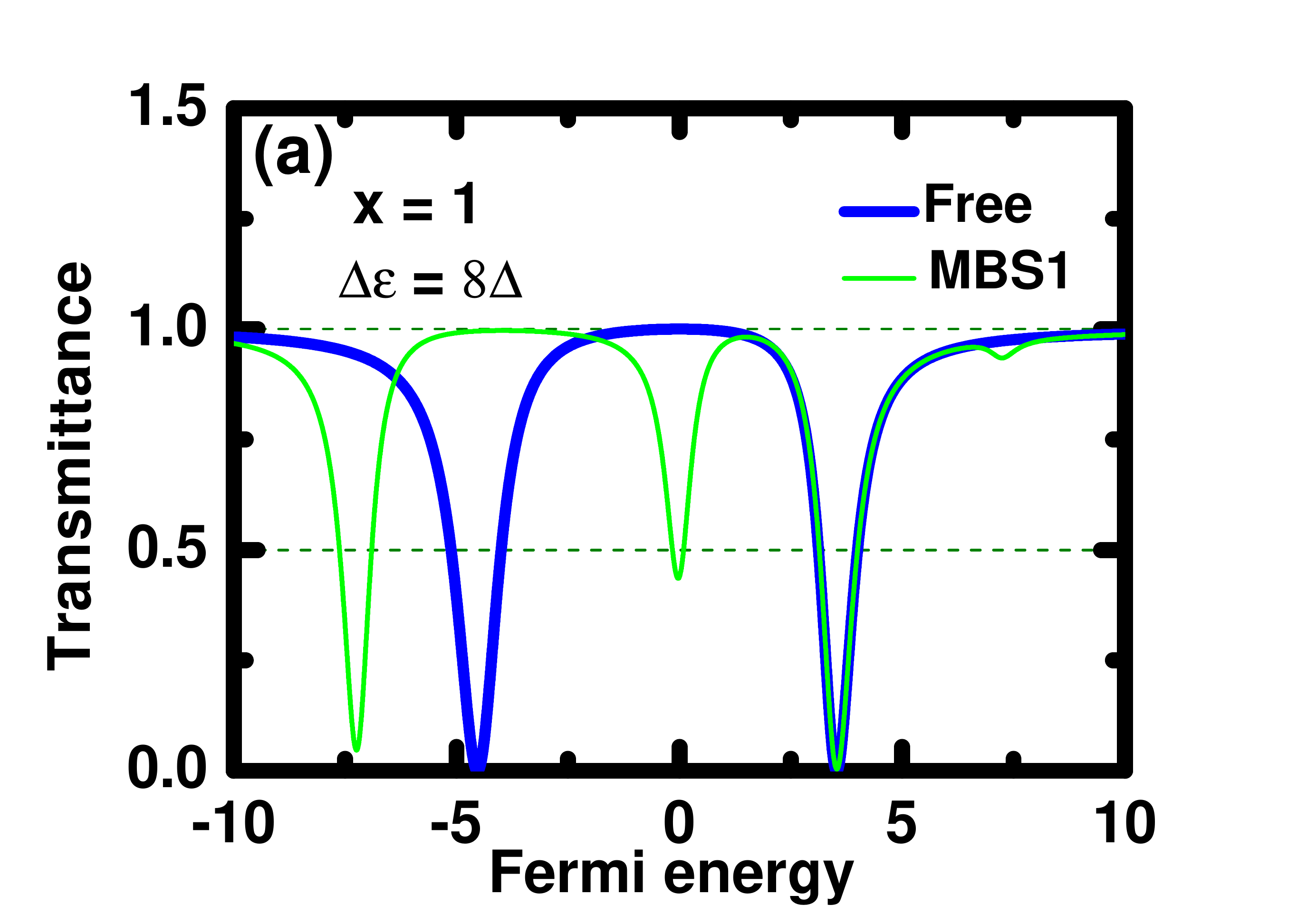}}}\centerline{\resizebox{3.3in}{!}{\includegraphics[clip,width=0.14\textwidth,height=0.06\textheight]{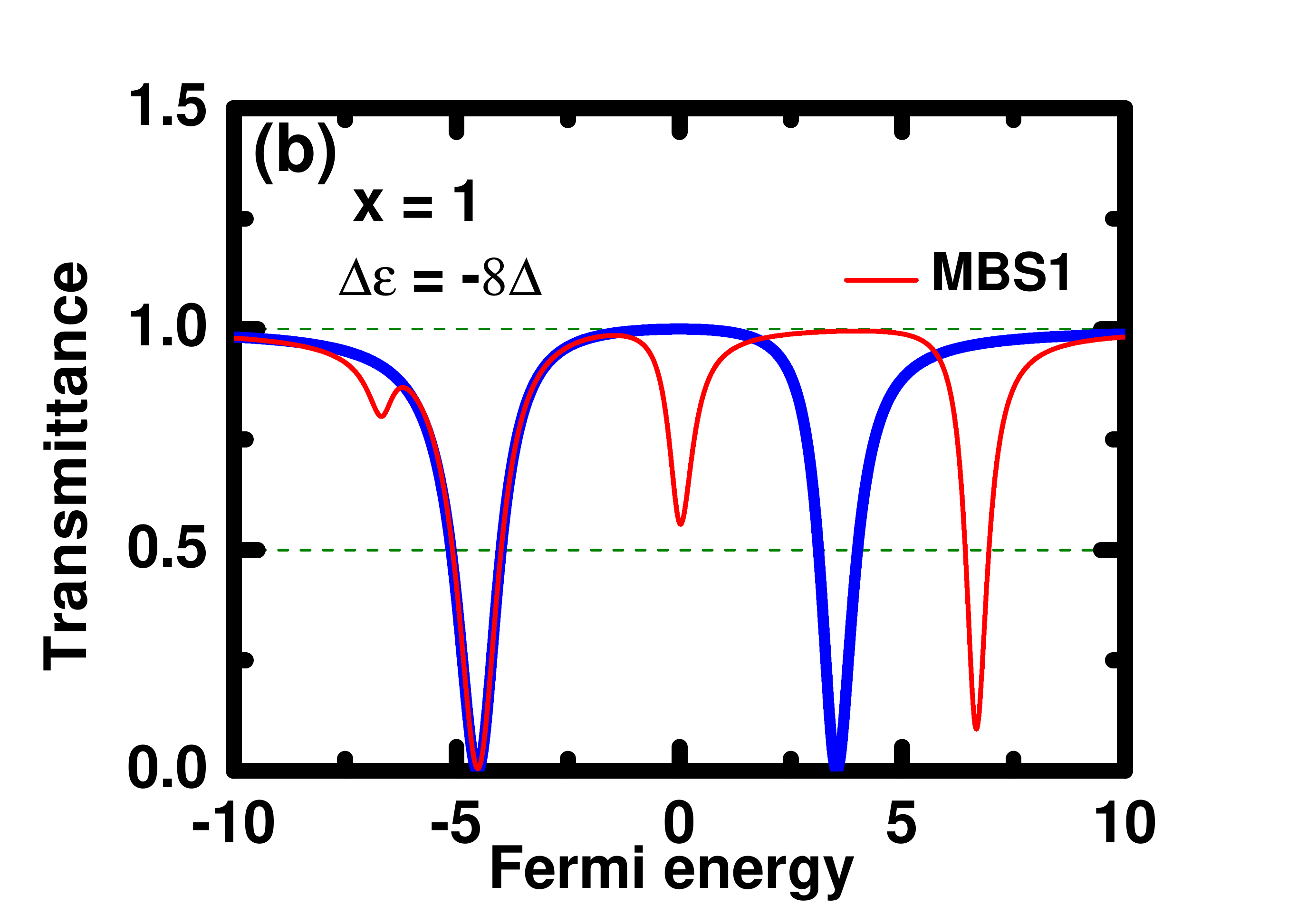}}}
\caption{(Color online) Transmittance determined by Eq.\,(\ref{eq:trans2}) as a function of the Fermi energy $\varepsilon$ in units of $\Delta$ in the Fano regime $x=1$ $(q_{b}=0)$ for the double QD setup of Fig.\,\ref{Fig1} with $\varepsilon_{M}=10^{-7}$. Panels (a) and (b): reversed patterns of panels (a) and (b) of Fig.\,\ref{Fig6}.}
\label{Fig7}
\end{figure}

\begin{figure}[!]
\centerline{\resizebox{3.3in}{!}{\includegraphics[clip,width=0.14\textwidth,height=0.06\textheight]{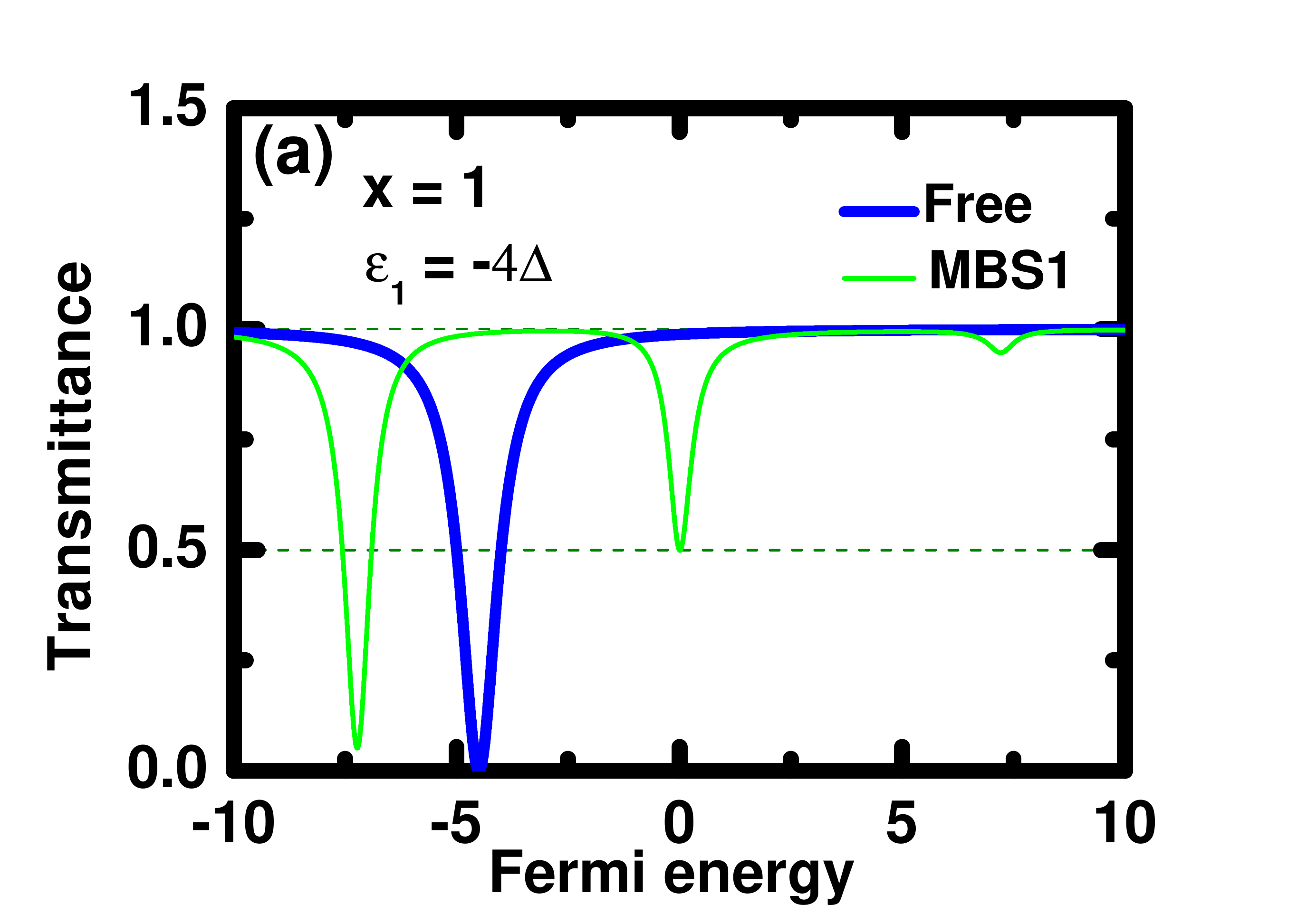}}}\centerline{\resizebox{3.3in}{!}{\includegraphics[clip,width=0.14\textwidth,height=0.06\textheight]{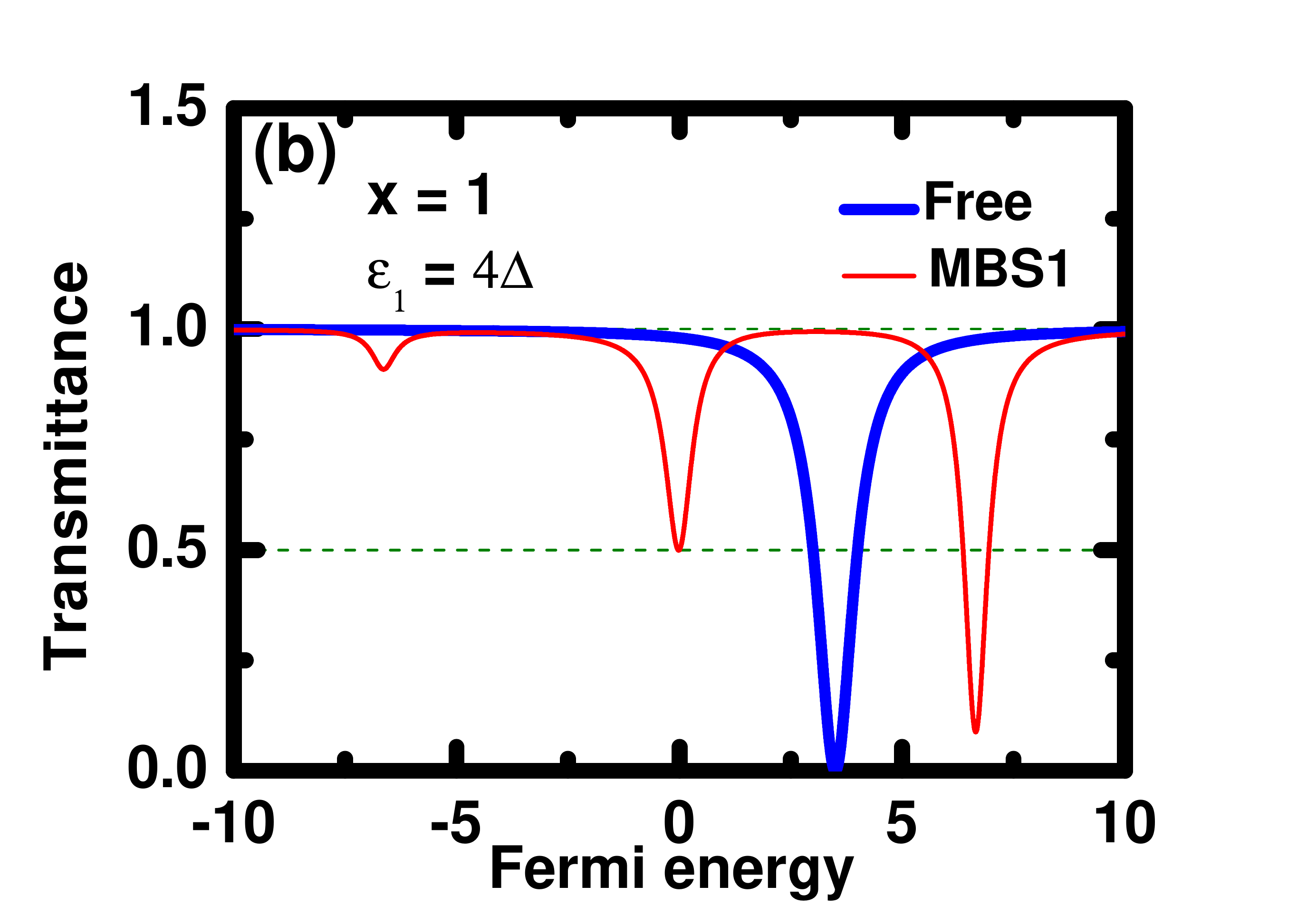}}}
\caption{(Color online) Transmittance determined by Eq.\,(\ref{eq:trans2}) as a function of the Fermi energy $\varepsilon$ in units of $\Delta$ in the Fano regime $x=1$ $(q_{b}=0)$ for the situation without the QD 2 in the setup of Fig.\,\ref{Fig1} by considering $\varepsilon_{M}=10^{-7}$. In panel (a) we use $\varepsilon_{1}=-4\Delta$: the solid-green lineshape is for the situation of the MBS 1 present, while the blue curve corresponds to the free interferometer in which the Kitaev wire is absent. Panel (b): For the reversed QD level $\varepsilon_{1}=4\Delta,$ the pronounced satellite dip of the QD $1$ in the free case (solid-blue lineshape) appears above the Fermi level of the leads as expected and the ZBA in the solid-red lineshape persists as observed in panel (a). Thus in both panels a Majorana dip (ZBA) emerges characterized by an amplitude of $1/2$.}
\label{Fig8}
\end{figure}

Fig.\,\ref{Fig7} contrasts the transmittance patterns of Fig.\,\ref{Fig6}. The Fano regime  $x=1$ $(q_{b}=0)$ replaces the resonances in Fig.\,\ref{Fig6} by the antiresonances of Fig.\,\ref{Fig7} as can be seen in panels (a) and (b) of the same figure. Thus the ZBA is a Majorana dip that drops to above or below $1/2$ depending on the sign of $\Delta\varepsilon.$ A detailed discussion on the mechanism of such a fluctuation is addressed in Ref. {[}\onlinecite{key-25c}{]} for an analogous Scanning Tunneling Microscope system.
To remove this instability of the Majorana fingerprint the QD $2$ should be discarded by keeping $\varepsilon_{M}=10^{-7}$ as shown in Fig.\,\ref{Fig8}(a) and (b): in the single QD setup the dip of the transmittance stabilizes at $1/2$ both for $\varepsilon_{1}=-4\Delta$ and $\varepsilon_{1}=4\Delta$. The independence of this hallmark with the gate voltage for a single QD also occurs in the opposite Fano limit $x=0$ $(q_{b}\rightarrow\infty)$ as widely treated in Ref. {[}\onlinecite{key-25b}{]}.

Additionally, we can conclude from Fig.\,\ref{Fig8} that the single QD setup avoids the manifestation of the symmetry-breaking feature in the system Green's functions, which are present in the double QD version of the interferometer as we have demonstrated in this work. Thus the methodology proposed here to pursuit Majorana bound states is not feasible by considering only one dot: the employment of a couple of QDs allows us to explore how the aforementioned symmetry is broken in distinct scenarios as those with a free interferometer, a side-coupled regular fermion and a Kitaev wire within the topological phase. Each one breaks the symmetry by its own way and the differences between these cases reveal the Majorana way of symmetry breaking: the signature of a Majorana is not restricted to the vicinity of the ZBA due to the zero mode nature of such an excitation, it spreads over the entire space spanned by $\varepsilon$ and $\Delta\varepsilon$ in the density plots of panels (b) for Figs.\,\ref{Fig2} and \ref{Fig3}. Panels (a) of Figs.\,\ref{Fig2} and \ref{Fig3} as well as Fig.\,\ref{Fig4} exhibit non Majorana fingerprints due to distinguished ways of breaking the symmetry property under consideration.

\section{Conclusions}
\label{sec4}

In summary, we have explored theoretically a semi-infinite Kitaev
wire within the topological phase coupled to a double QD system. Our
analysis  reveal that the Green's functions of the QDs are antisymmetric
under the permutation of the indexes that label the parameters of
these dots. We propose that such a feature can be probed
experimentally just by measuring the zero-bias conductance as a function
of the Fermi energy of the leads and the symmetric detuning of
the QDs levels.

Thereby, our results reveal a strong dependence of
the transmittance on contrasting Fano regimes of interference, where
conducting and insulating regions emerge as signatures of a single
Majorana fermion excitation. We have demonstrated that the symmetric Fano
interference is recovered in particular for the electron tunneling
only via the QDs and with the metallic leads in resonance with the
Majorana zero mode, thus resulting in a fluctuation of the transmittance given by half or unity,
respectively for a semi-infinite or finite Kitaev wire. The former
situation corresponds to an isolated Majorana excitation, while the
latter represents the bounding of two distant Majoranas and the formation
of a nonlocal Dirac fermion. Both scenarios can be helpful to guide experiments whose purpose is to pursuit the existence of a single Majorana fermion or a pair formed by such a quasiparticle in condensed matter systems.

\begin{acknowledgments} This work was supported by the Brazilian
agencies CNPq, CAPES, FAPEMIG, PROPe/UNESP and 2014/14143-0 São Paulo Research Foundation (FAPESP). A. C. Seridonio and F. A. Dessotti are grateful to P. Sodano for valuable discussions.
\end{acknowledgments}


\begin{thebibliography}{10}
\bibitem{OBS} The current lack of symmetry in the system Green's functions was verified previously in an STM system with adatoms \cite{key-25c},
but for a Fermi energy in resonance with the Majorana zero mode. In the case of a setup composed by QDs as the proposed in this manuscript, the Fermi level is an experimentally tunnable parameter thus allowing the feasibility of a situation away from the resonance condition.
\end{thebibliography}
\end{document}